%% file: main.tex
\documentclass[journal,10pt]{IEEEtran}
\usepackage[utf8]{inputenc}
\usepackage{hyperref}
\title{Landing AI on Networks: An equipment vendor viewpoint on Autonomous Driving Networks}
\author{Dario Rossi$^1$, Liang Zhang$^2$\\
\emph{$^1$Huawei Technologies, co. Ltd, Paris Research Center} --- \texttt{dario.rossi@huawei.com}\\
\emph{$^2$Huawei Technologies, co. Ltd, Nanjing Research and Development Center} --- \texttt{zhangliang1@huawei.com}\\
}
\date{Jan 2022}

\usepackage{booktabs}
\usepackage{cite}
\usepackage{tabularx}
\usepackage{comment}
\usepackage{url}
\usepackage{graphicx}
\usepackage{color}
\newcommand{\DR}[1]{\textcolor{red}{#1}}
\newcommand{\censure}[1]{}

\newcommand{\firstfakepar}[1]{\noindent\textbf{#1}.}
\newcommand{\fakepar}[1]{~\\\noindent\textbf{#1}.}

\begin{document}

\maketitle

%%%% 1auth et al.
\bstctlcite{IEEEexample:BSTcontrol}

\thispagestyle{empty}
\begin{abstract}
The tremendous achievements  of  Artificial Intelligence (AI) in  computer vision, natural language processing, games and robotics,  has extended the reach of the AI hype to other fields: in telecommunication networks, the long term vision is to let AI fully  manage, and autonomously drive, all aspects of network operation. 
In this industry vision paper, we  discuss challenges and opportunities of Autonomous Driving Network (ADN) driven by AI technologies.  To understand how AI can be successfully landed in current and future networks, we start by outlining  challenges that are specific to the networking domain, putting  them  in perspective with advances that AI has achieved in other fields. We then present a system view, clarifying  how AI can be fitted in the network architecture.  We finally discuss current  achievements as well as future promises of AI in networks, mentioning a roadmap to avoid bumps in the road that leads to true large-scale deployment of AI technologies in networks.
\end{abstract}


\section{The New Gold}

\IEEEPARstart{T}{he} last decade has witnessed significant advances in several fields where Artificial Intelligence (AI) has been applied to -- from image recognition, to natural language processing and  gaming  to name a few examples.
Such achievements are due to the fortunate confluence of several necessary ingredients: namely, (i) exceptional theoretical advances in the last 50 years, coupled to the availability of (ii) massive volumes of data, and equivalently (iii) massive computing capabilities. These achievements have gained significant press attention, fueling the hype on  AI techniques, and their expected benefits.
As a result, every technology sector joined this new ``Gold rush'', including   the networking field, where AI is envisioned on the long-run to fully  manage, and autonomously drive, all aspects of network operation~\cite{self-driving-net}.

At the same time, a significant fraction of AI 
projects are difficult to transfer\footnote{As Gartner put it\cite{80fail}, in 2020 ``80\% of AI projects will remain alchemy, run by wizards whose talents will not scale in the organization''} beyond the initial proof-of-concept.
The difficulty in successfully landing AI is overtly recognized lately~\cite{oreilly}, and is technically rooted in either the lack of some of the above ingredients, difficulty in integration, or other non-technical aspects\censure{(such as company-level management aspects including culture or talent that we will not directly discuss here)}. As ``not all gold that glitters,''  it is necessary to understand what type of problems AI can solve, and how AI solutions can be fit in the overall system: this is necessary, in order for AI to really make the difference in a specific technology field, such as the networking domain considered in this paper, before the next AI winter.

A decade ago, Marc Andreessen was rightly anticipating that ``software is eating the world'': in the last decade, evolution toward software enabled networking world to escape ossification~\cite{anderson2005ossification}, and AI-software shows   the very same appetite for the next decade.   Owing to growing success of  all-IP (2000-2010) and cloud-native (2010-2020) networking technologies, IP-enabled communications are now spanning a very large (and still growing) set of vertical sectors and markets. To manage such a plethora of heterogeneous services evolving at a  fast pace,  the network operation and management (O\&M) community has started turning its attention to  AI, for relieving and assisting human operation for diverse tasks (e.g., ranging from configuration, to dynamic resource management, troubleshooting and quality assessment). 
In a field where a significant fraction of the operations are still involving human intervention, and where such interventions are also responsible for a significant fraction of the errors, AI seems an appealing means to immediately automate part of these manual tasks (e.g.,  from configuration, to fine-grained resource management at very fast timescale, troubleshooting guidance and quality forecast)  and later  reach fully automated and error-free (or at least self-healing) operations. 

Clearly,  the evolution of the network O\&M to 
a fully autonomous driving  network (ADN), cannot be done overnight due 
to technical challenges, adoption barriers and legal aspects (e.g., liability).
Pragmatically, our vision is thus for AI to replace human hands in the  \emph{fast loop}, but not fully supplant humans which is
essential to keep in the \emph{slow loop}. Taking the viewpoint of an equipment vendor, this paper illustrates the current status of network AI, enriching the narrative with  examples of research that successfully landed in network technologies, further  laying out the steps necessary to the ADN for reaching  higher level of autonomicity and intelligence.  We instead disregard complementary technological aspects  (well covered by~\cite{jacquenet2021viewpoint,wiley1}) and methodological aspects (for which we refer the reader to \cite{boutaba,tnsm1,jsac1,jsac2,wiley1}).

%\DR{At the same time, the network evolution from a set of heterogeneous protocols, governed with
%policies, configurations and heuristics  INTRISE of  intensive human expertise to a fully autonomous self-governing network cannot be done overnight. The problem is not only technical,
%which would anyway be challenging per se, but also of adoption -- see e.g., the field of  autonomous driving \emph{cars}, where part of the ethical debate would remain even in case  where autonomous driving cars were to be significantly safer than human drivers.Clearly why ethics may be more difficultly cast to the field o communication networks,  where wrong decisions  are less likely to cost human lifes, the fact of yielding control  to a completely different set of algorithms,  is far from beign trivial. }.

%footnote{80\%
%https://blogs.gartner.com/andrew_white/2019/01/03/our-top-data-and-analytics-predicts-for-2019/}

The rest of this paper is organized as follows.
Sec.~\ref{sec:ai} briefly introduces AI, and Sec.~\ref{sec:cfr} overviews AI challenges on the network domain, putting them in perspective against other fields where AI has been successful. 
Sec.~\ref{sec:adn} then introduces the key aspect of the Autonomous Driving Network (ADN), examine its architectural, hardware and software needs  with a focus on AI-related aspects. Finally,  Sec.~\ref{sec:current}  overviews illustrative examples of how AI can be successfully landed in current networks,  while  Sec.~\ref{sec:future} discusses open and future challenges on the path towards the ADN.

%\tableofcontents

\section{What is AI, and why it matters ? }\label{sec:ai}
\input{10_ai}

\section{Landing AI on Networks }\label{sec:cfr}
\input{20_cfr_alternate}

\section{Anatomy of an Autonomous Driving Network}\label{sec:adn}
\input{30_adn}

\section{Current achievements of network AI}\label{sec:current}

\input{40_current}

\section{Future challenges for network AI}\label{sec:future}
\input{50_future}

\section*{Acknolwedgements} 
The authors wish to thank Filip De Turck and Prosper Chemouil
for their feedback and useful discussion on the initial versions of this manuscript, as well as the IEEE TNSM EiC Hanan Lutfiyya and the anonymous Reviewers,  whose useful comments helped improving the quality of the final manuscript.

\bibliographystyle{IEEEtran}
\bibliography{references}

\begin{IEEEbiography}[{\includegraphics[width=1in,height=1.25in,clip]{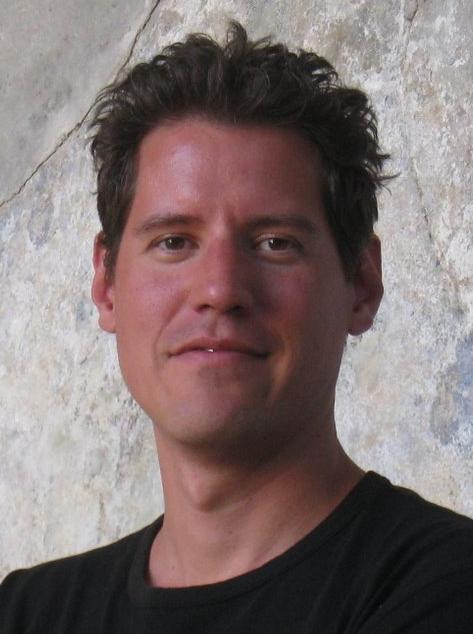}}]{Dario Rossi}
is Director of Huawei AI4NET Lab and  Director of the DataCom Department at the Paris Research Center, France. Before joining Huawei in 2018, he held Full Professor positions at Telecom Paris and Ecole Polytechnique and was holder of Cisco's Chair NewNet\@ Paris. He has coauthored 15 patents and over 200 papers in leading conferences and journals, that received 9 best paper awards, a Google Faculty Research Award (2015) and an IRTF Applied Network Research Prize (2016). He is a Senior Member of IEEE and ACM.
\end{IEEEbiography}

\begin{IEEEbiography}[{\includegraphics[width=1in,height=1.25in,clip]{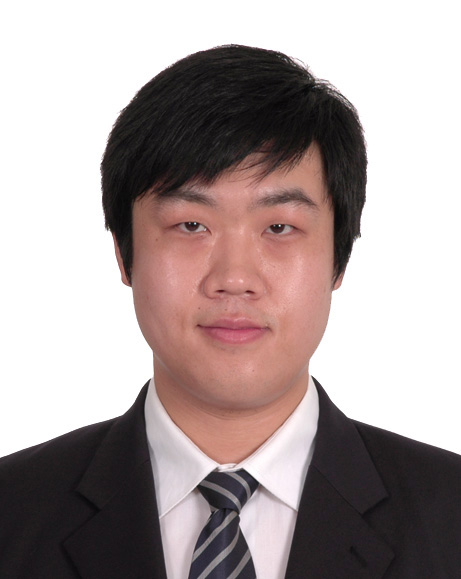}}]{Liang Zhang}
is Vice-Director of Huawei AI4NET Lab and  Director of the DataCom AI Department at the Nanjing Research Center, China.  
He received the PhD degree from Southeast University, Nanjing, China, in 2010.  His research interests include intelligent fault analysis, network traffic analysis and network optimization.

\end{IEEEbiography}

\end{document}

%% file: 10_ai.tex
As AI has recently become an abused term, and given that the very same definition of AI is constantly redefined\footnote{Due to the well known ``AI effect'', as soon as AI successfully solves a (narrow) problem, the problem is no longer considered a part of (general) AI.}, we briefly introduce it here what we consider to be the set of AI techniques that will constitute ADN's basic building blocks.

\subsection{Brief history of AI \& ML}
As visually depicted in the top part of Fig.\ref{fig:ai:intro}, the history of AI \& ML can be traced traced back to  Alang Turing's work in the early 1950s\censure{1950,
Alan Turing ``imitation games''}  with the
basis of Neural Networks (NN), as well as the terms Artificial Intelligence (AI)
and Machine Learning (ML) introduced during that decade\footnote{AI was first mentioned in the 1955 Dartmouth Summer Research 
project co-led by John McCarthy; the  basic building blocks of NN, i.e., the Perceptron, was introduced in 1957\cite{rosenblatt57perceptron};  ML was first mentioned in 1959 by Arthur Samuel related to 
a computer program learning to play checkers\cite{samuel59ml}}.
Clearly, AI/ML techniques have evolved significantly since then: as for many scientific fields, their evolution has been shaped by ``hype'' cycles, where peaks of attention and spending sprees (known as ``AI springs'') were followed by periods of disengagement and funding cuts (known as ``AI winters'', in the late 70s and late 80s). 
While the network community pioneered the use of AI as early as in the 70s\cite{narendra1977application}, many of the now popular branches of AI,  such as  Deep Learning (DL) and Data Mining were introduced in the last AI spring\footnote{Deep learning (DL) was first mentioned in 1986 by  Rina Dechter and Knowledge discovery in databases (KDD) in 1989 by Gregory Piatetsky-Shapiro; interestingly, it is not until the early 2000 that DL was used in association with NN, while  the first breakthrough of Deep Neural Networks (DNN)  in the field of computer vision\cite{alexnet2012nips} had to wait for another decade.}.

%\DR{in the rest of this section we pinpoint relevant AI techniques and example of successful  application, extract the main reasons of their success.}
%For instance times of the  Perceptron\footnote{rosenblatt57perceptron}, the basic building blocks of Artificial Neural Network (ANN)
%\DR{out of place:  more recent discoveries that biological neurons are actually much more complex (A 5-8 layers artificial deep neural network  is necessary to model a single neuron \cite{beniaguev2021neuron} than artificial ones)}.

\subsection{Crude taxonomy of AI \& ML}
\begin{figure}
    \centering
    \includegraphics[width=0.85\columnwidth]{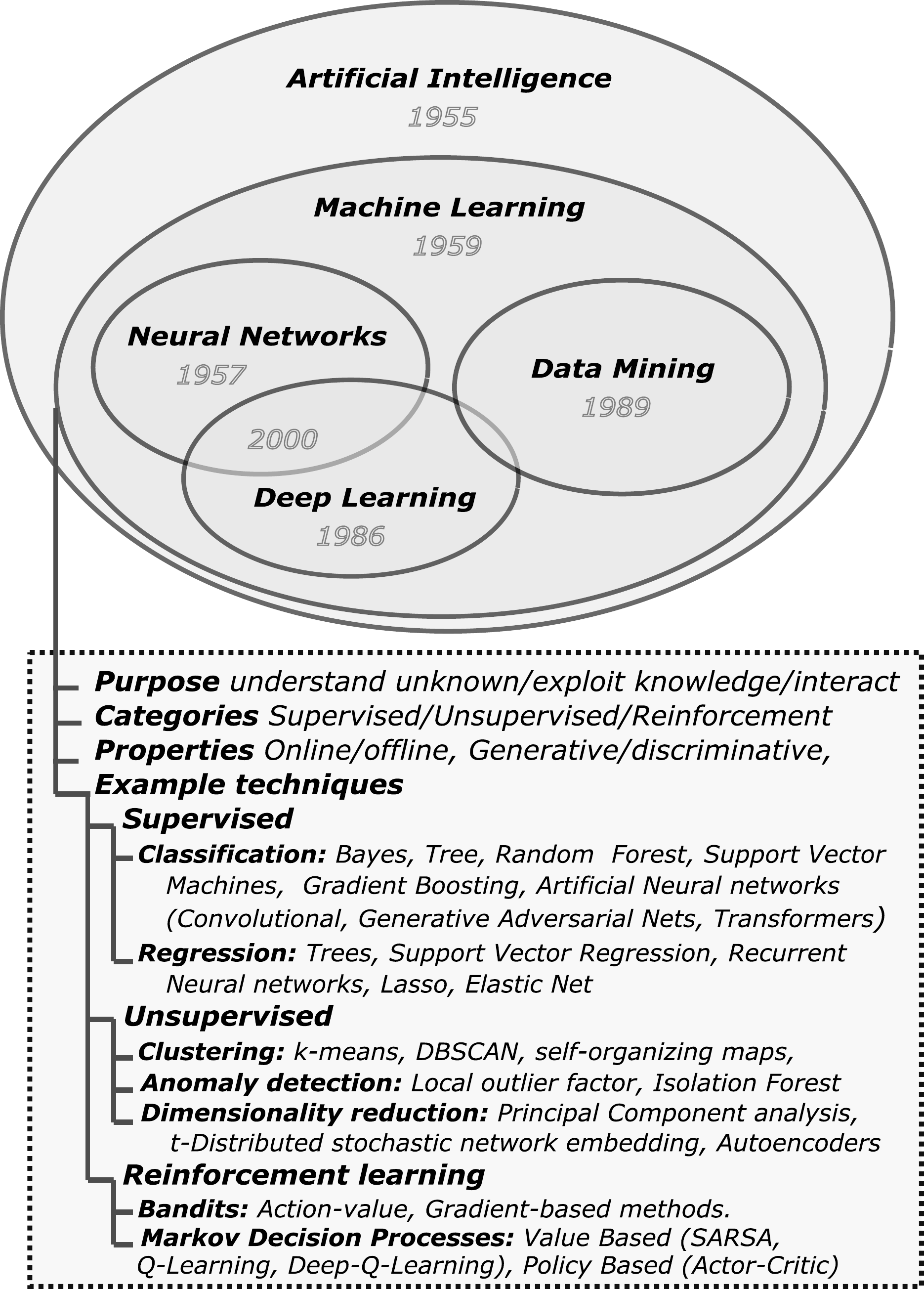}
    \caption{Brief history and crude taxonomy of AI.}
    \label{fig:ai:intro}
\end{figure}

As AI is again reaching very high hype levels, it is a valid question to ask which AI techniques, and to what extent,  can be successfully landed in the network field before the next ``ice age'' of AI.
In the context of this paper, we are not interested in establishing a rigorous taxonomy of AI \& ML techniques.
Still, given that these terms have been abused up to being the target of popular jokes~\cite{matvelloso18tweet_mlvsai}, it is useful for clarity (and simplicity) to refer to a (crude) taxonomy reported in the bottom of Fig.~\ref{fig:ai:intro}.
For the scope of this paper, we restrain the focus to techniques that belong to  the data-driven branch of AI that is also known as ML. By abuse of language, we will use the two acronyms interchangeably in the following.  

%, nor we are interested in the broad emergence of general AI: rather, we are interested in discussing how the set of techniques that undergo the AI \& ML umbrella can be succesfully exploited in a networking context.

From a very high level, the purpose of ML can be either to
(i) make effective use of \emph{existing} knowledge, or (ii) gather structured understanding on \emph{unknown} phenomena, as well as (iii)  \emph{learn} to achieve a goal. Roughly, these purposes directly map to three branches of ML that are respectively known as  (i) supervised,  (ii) unsupervised  and (iii) reinforcement learning -- though  blending among categories is possible (e.g., semi-supervised or self-supervised learning).  While a more detailed and comprehensive taxonomy  is out of scope, examples of relevant techniques for each branch are provided in the bottom of Fig.~\ref{fig:ai:intro}. Oversimplifying, many (though not all) of the ML techniques are efficient ways to solve complex data-driven optimization problems, offering solutions that are well suited for the data at hand (i.e., fit well), but are also amenable to generalization (i.e., avoid overfitting).  This constitutes part of the reasons of ML success in several fields, and  makes it practically appealing for the network domain too.

\subsection{Example of AI success across all domains}
The most recently hyped examples of  success in the current ``AI spring''  pertain to areas such as computer vision,  game-playing and  natural language processing.  

%Other areas equally benefited from AI techniques,  such as e.g., computational biology (protein unfolding), recommendation system (Netflix, advertisement business), which may be  less hyped though with respect to the former.
Image recognition attracted significant attention not only as being among the first key success of Convolutional Neural Networks (CNN)~\cite{alexnet2012nips}, but also e.g.,  in reason of the powerful  Generative Adversarial Networks (GAN)~\cite{goodfellow2014gan} underneath the (in)famous DeepFakes technologies~\cite{deepfake2018obama}.
In the game-playing context, Deep Reinforcement Learning (DRL) has been instrumental in achieving super-human playing abilities, with e.g., Google's AlphaGo\cite{alphago} beating the go board game world champion Lee Sedol, or OpenAI Five\cite{openaiFive} winning the online computer-game DOTA2 tournament.
In natural language processing, self-supervised neural embedding (e.g., \emph{word2vec}
\cite{mikolov2013word2vec}) and  few-shot transformer technologies
\cite{brown2020arxiv} such as OpenAI GPT3\cite{gpt3} gained significant traction lately. 

As we illustrate in Sec.\ref{sec:current}, communications and computer networks are one among the numerous other domain of applications (biology, medical field, robotics) that is currently exploring the  use of AI techniques in many aspects of its operation.

%% file: 20_cfr_alternate.tex
\newcommand{\landingai}[0]{Landing AI in Networks}

To replicate AI success in  other fields, it is necessary to first understand their root cause: we thus start by dissecting AI successes, 
to draw informed conclusions from a networking perspective\footnote{This paper adopts an ``AI for networks'' angle, i.e.,  how  AI can improve network Operation \& Management (O\&M).  We point out that, in reason of  AI success, an equally important viewpoint that research has considered is ``networks for AI'', i.e., how networking techniques can make AI workflow more efficient, where AI becomes thus a networked application -- which we instead disregard due to space constraints.}  for the ADN.

Without willing to undermine the significance of AI achievements, we
remark that recent advances have equally benefited from:
(i)  ground-breaking advances in ML theoretic research,
(ii) the availability of large corpora of (labeled) data to feed ML models with, and  (iii) the availability  of software platforms that efficiently exploit hardware acceleration.  
Two observations are worth sharing: on the one hand,  whereas the latest wave of AI success essentially involves neural networks, this is clearly not the first time that AI accomplishes similar prowess; on the other hand, this AI spring may be the first time at which time is ripe for advances in all (i), (ii) and (iii) aspects.  Aside from the above technical aspects,  success  also  depends on (iv) business considerations. 
These four aspects are crucial for landing AI in networks, as we next discuss.

% Previous
%success stories in the very same fields of image recognition, 
%game-playing and language processing, include breakthrough in 
% Optical Character Recognition (OCR) by Support Vector Machines 
% -- it is sufficient to recall that IBM DeepBlue defeated Chess world champion Garry Kasparov in 1997\cite{deepblue}, 
% the CVPR prowess in Optical Character Recognition (OCR)  
% in the XXXX (SVM)
% or for NLP interaction (chatbots).
% However the theoretical advances have been 

\subsection{Theoretical advances} 
\firstfakepar{Root of AI success} The latest AI spring yielded to numerous theoretic advances. The most visible ones, in reason of their recent hype,  involve theoretic advances of Deep Neural Networks~\cite{dnnbook}, for which  Y.\,Bengio, Y.\,LeCun and G.\,Hinton were recently  distinguished with the 2018 ACM Turing award. 
Other equally important advances include 
such as the ensembling of several ``weak'' classifiers -- that became widely known after the popular Netflix-prize~\cite{netflix-prize}. Ensembles are now state of the art for many  supervised\cite{breiman2001randomforest,chen2016xgboost} and  unsupervised \cite{liu2008isolationforest} tasks. Similarly, advances in causal reasoning capabilities\cite{pearl2019cacm}  for which J. Pearl has been credited the 2011 ACM Turing award, are perhaps less known to the broad public, but equally relevant. Finally, more prospective advances (e.g., spiking neural networks~\cite{spiking}) are under way, but are in still early stage of development so they can be expected to have a deeper impact on a longer time horizon.

% \DR{
% One appealing aspect of NN is to move away from "Ferature enginerring".
% At the same time, the NN is architecture over-engineering }

\fakepar{\landingai}  Fortunately, the  greatest inventions\footnote{The foundation of the TCP/IP architectural principles of the Internet is traced back to mid 70s, for which  V.\,Cerf and B.\,Kahn were credited with the 2004 ACM Turing award.
Similarly, the foundation of Internet's most successful application, i.e., the World Wide Web, is traced back in the late 80s, for which T.\,Berners-Lee received the 2017 ACM Turing award.}
of our time has democratized access to knowledge: as such, all AI  advances can be readily used in the network field.
Clearly, whereas many of these recently hyped successes involve DL technologies,  it appears evident that also any of the lesser-hyper AI technologies reported  in Fig.\ref{fig:ai:intro} is still worth considering from the viewpoint of an AI-fueled autonomous driving network.  
Purposely, we later (see Sec.\ref{sec:current}) report on successful network O\&M application using AI techniques spanning all AI branches.

\subsection{Hardware and software}\label{sec:cfr:hwsw}
\firstfakepar{Root of AI success} Ultimately, theoretical advances are  distilled in algorithms, for which hardware and software engines running them are equally important.  From a \emph{hardware} viewpoint, DNN have benefited from  the emergence and commoditization of hardware acceleration such as  Graphic Processing Units (GPUs), that are widely acknowledged to have substantially contributed to the recent success of AI\cite{hwlotteryCACM21}.
A further evolution is the emergence of domain-specific architectures (DSA), that are well discussed by J.\,L.\,Hennesey and D. Patterson, in their 2017 ACM Turing award inaugural lecture~\cite{turingAward2017cacm}: Tab.~\ref{tab:tpu} reports examples of DSAs for  NN acceleration.  
Worth mentioning are  advances on binary~\cite{hubara2016binarized} and xor~\cite{rastegari2016xnor} neural networks, as well as neuromorphic processors~\cite{davies2021neuromorphic} for  spiking networks acceleration, of  interest on a longer time horizon.

The availability of \emph{software} stacks able to efficiently leverage the above hardware in a seamless manner, jointly providing a unified and complete environment is key to lower bootstrap cost of AI in any field, as well as to facilitate transfer. Popular stacks include  Google TensorFlow\cite{tensorflow}, Huawei MindSpore\cite{mindspore}, Telsa Pytorch\cite{pytorch} for neural network, and  Scikit-learn\cite{scikitlearn}, for general workflows.  These stacks offer the ability to rapidly prototype in high-level language, with Python commonplace now in the  scientific workflow, and have optimized backends seamlessly supporting hardware acceleration. Software stacks have been at least as important as hardware accelerators (if not more according to \cite{declarativeMLCACM22}) to lower  AI entering barrier, with respect to the more scattered situation of just less than a decade ago.

%33.90GFLOPS = 0.03

\begin{table}[t]
\caption{Generic CPU, GPU vs Domain-specific hardware accelerators}\label{tab:tpu}
\begin{tabular}{lllp{14pt}ll}
\toprule
\bf Vendor & \bf  Product   & \bf Target &  \multicolumn{2}{l}{\bf Processing} & \bf Power  \\
       &           &         & {\scriptsize [TOPS]} & {\scriptsize [TFLOPS]}     & [W] \\
\midrule
ARM    & Cortex A72 \cite{cortexA72} & Edge CPU   &     n.a. &  0.03      &  0.75\\
Google & Coral.AI  \cite{coral.ai}  & Edge  DSA  &     4 &  n.a.    &  2 \\
Huawei & Ascend310 \cite{ascend310} & Edge  DSA  &    22 & 11   &  8 \\
\midrule
Intel  & Xeon 8280    \cite{intelXeon8280} & Cloud CPU &  n.a.       &  $\approx 2$  & 205     \\
NVIDIA & P100       \cite{nvidiaP100} & Cloud GPU &  n.a.  & 5-21    & 250  \\
Google & TPUv3     \cite{tpuv3}     & Cloud DSA  &   n.a.  &  420 & $\approx$ 300   \\
Huawei & Ascend910 \cite{ascend910} & Cloud DSA  &   640  &  320 & 310\\
\bottomrule
\end{tabular}
\end{table}

%\footnote{\url{https://deepmind.com/research/case-studies/alphago-the-story-so-far}}
% \footnote{\url{https://openai.com/projects/five/}}.

%DeepBlue attained super-human playing capabilities essentially via unprecedented search capabilities~\cite{deepblue}, resulting from a single-chip chess search engine with massive  parallelism (capable of scanning 700,000 positions per second),  a complex evaluation function (relying on 8000+ features to assess the move quality), and effective use of a Grandmaster game database (essentially, a book of 4000 openings hand-coded by chess Grandmaster Joel Benjamin that acted as consultant to the DeepBlue team).

%AlphaGo (10,000s of human amateur and professional games, 3 days training, 1920 CPUs, 280 %GPUs, elo rating 3.16) 
% AlphaGo Zero (simply plays against itself)  4 TPUs, 40 days to beat AlphaGo Master, %achieving elo 
% Our Dota 2 AI, called OpenAI Five, learned by playing over 10,000 years of games against itself. 

\fakepar{\landingai}  The set of hardware accelerators and software stacks just introduced for the general AI case,  undoubtedly facilitates development and execution of AI workflow for network as well.
At the same time, whereas Cloud-native workflows can expect to find GPUs or high-end TPUs of Tab.\ref{tab:tpu}, many of the network operations will need to be carried out without having access to cloud resources. Similarly, whereas the factor form and their power drain of some TPU chipsets (eg. Ascend310 or Coral.AI) are small, however this does increase Capex (for the new chip) and Opex (higher computational cost, albeit small, for the new AI feature). As such, whereas SmartNICs equipped with such powerful AI accelerators start to appear~\cite{nvidiaEGX}, it is difficult to predict yet how much they will be widespread in networking equipments. Thus, a safer bet is to consider an AI workflow that can exploit AI acceleration if available, but is lean enough to run on standard CPUs (e.g., as for the case of ARM, using software acceleration libraries such as armNN\cite{armNN}, that can seamlessly support Cortex-A CPUs and Mali GPUs).

Additionally, to optimize communication between CPU and the GPU/TPU boards/chipsets,  GPU and TPUs are designed for batch processing, with furthermore relatively large batch size of several thousands elements -- which is in contrast with the needs of typical network workflow. Indeed, while networks stack employ batch processing for \emph{packet-level} operation~\cite{PROCIEEE-19},  the size of batches is one to two orders of magnitude smaller. Additionally, in typical networking use-cases, most AI processing should happen instantaneously, i.e., with batch size equal to 1. Considering for instance \emph{per-flow} decisions, it appear  inconvenient to batch AI processing across multiple flows: e.g., waiting for the arrival of several flows   due to batching would  require buffering traffic, couple decisions and delay application of AI-driven policies.  Fortunately,  the emergence of stacks such as TensorFlow-Lite (TFL)\cite{tflite}, or more recently TensorFlow-Lite micro (TFLM)\cite{david2021tensorflow-lite_micro}, makes it easier to run proof-of-concept AI models on  constrained devices and CPUs  that are pervasive in network equipment.

Shortly, we argue that while hardware accelerations starts being available, due to the KISS principle, AI solutions that are excessively computationally costly are not going to be successfully landed in networks -- an aspect worth further attention that we thus consider in the following discussion.

\subsection{Data (and environment)}\label{sec:cfr:data}
\firstfakepar{Root of AI success}
The third key of success has been the availability of large datasets (or controlled environments). For instance, MNIST~\cite{mnist}, CIFAR~\cite{cifar} or ImageNet~\cite{imagenet}, that overall comprise tens of millions of images for tens of thousands of classes,   have been instrumental to fuel advances in image recognition~\cite{alexnet2012nips}.
Similarly, recently hyped advances~\cite{gpt3} in NLP relied on
hundred billions of text tokens corpus  such as Common Crawl~\cite{commoncrawl}. Even in lesser mass-mediatized fields such as computational biology, it is fairly well recognized that astonishing advances\censure{AlphaFold-2}~\cite{alphafold-2} would not have been possible without 50 years of expert-driven labeling work on protein unfolding.
%%%%
In the reinforcement learning branch of AI, one or more agents interact with an environment to learn a successful strategy, by  enforcing actions that alter environmental responses. In this playground, OpenAI Five~\cite{openaiFive} learned by playing over 10,000 years worth of games against itself and  AlphaGo-Zero~\cite{alphago} was trained with  29 million games of self-play during  40 days using 4\,TPUs. Clearly, the ability to super-scale the  exploration of the action space in a faster-than-real-time yet realistic-enough environment has been crucial to achieve such results.

\fakepar{\landingai} Unconstrained access to high quality data, which is key for accurate models with good generalization capabilities, is a known problem across all AI domains of applications~\cite{noaiwothoutdata2011cacm}, and networking is not an exception. Interaction with an environment in a closed  learning loop  further exacerbates the problem.
 
Considering the classic ``4V'' data properties for the sake of simplicity, in terms of \emph{volume} and \emph{velocity}, it is sufficient to recall that the data rate of a single ToR switch 
is higher than the high-volume physics collected by Stanfords' Large Synoptic Survey Telescope~\cite{LargeSynopticSurveyTelescope}
or CERN's Large Hadron Collider (LHC)~\cite{LargeHadronCollider}. Additionally, if  the widely popular Moore law postulated a exponential growth in the computing capacity, the lesser known but equally important Gilder law observes that {``Bandwidth grows at least three times faster than the compute power''}, i.e., making the matter worse than in other fields. In terms of \emph{variety}, network data is extremely scattered, multi-modal, heterogeneous, more than what typically happens in domains such as image or natural language processing. Therefore, a standard and homogeneous data representation  would significantly facilitate AI application in networks. Furthermore, such heterogeneity make AI generalization capabilities of paramount importance.

Finally, in terms of \emph{veracity}, it appears evident that whereas other fields have crowdsourced, amassed and shared large datasets, the networking field is lagging far behind. This is due legal/business constraints on data sharing on the one hand (see Sec.\ref{sec:future:data}), and on the intrinsic difficulty of the labeling task on the other hand.
For instance, for image or NLP use-cases industries are either crowdsourcing labeling to  the whole population of Web users  (e.g., where every day 
an estimated 500 years of manpower~\cite{captchamadness}  is used to solve CAPTCHA puzzles to identify buses, cars and pedestrian in images, for training the future generation of self-driving vehicles), or directly recruiting  human labor  (e.g., speech recognition for digital personal assistants employs significant amount of human labor~\cite{anatomyofai} at all steps, including to sanitize, verify and validate transcriptions).  Conversely, labeling of, e.g.,  network anomalies, or identification of their root cause, or encrypted application traffic, is significantly more difficult to crowdsource as it requires domain expertise and large amount of time of skilled workers, which makes labeling cost higher.

On the one hand, we remark that a number of techniques  and good practices can help with \emph{data-related} (e.g., few-shot~\cite{few-shot-survey} or active~\cite{active-learning} learning techniques both help reducing the number of labels) and \emph{environment-related}  (pre-train models for transfer learning and fine-tuning in the real-environment~\cite{ecn2021sigcomm})  problems.  On the other hand, we stress that AI solutions that are unable to be continuously updated and fail to seamlessly generalize are not going to be successfully landed in networks either -- a  second aspect worth to systematically consider in the remainder of this paper. 

% Cite 80\% of time spent on data; Cite CEO surverys.

\subsection{Business (and beyond)}\label{sec:cfr:business}
\firstfakepar{Root of AI success}   Concurring in  the ultimate success of a technology are also non-technical aspects, which are at least worth  mentioning. For instance, for AI to be successful, it needs to solve an open (or otherwise unsolvable) problem, or solve it in a (significantly) more cost-effective way. By  rephrasing Hockam razor, we may consider that if simpler solutions are available  yielding good enough results at a fraction of the cost, then the simpler solution will be adopted (e.g., as it happened for the winner solution of the Netflix prize\cite{netflix-prize}, that was finally not deployed). 
Organizational aspects are  complementary and equally relevant: these include, the relevance of the selected business case, the
availability of AI profiles (data scientist, data engineers), and the culture of the enterprise (e.g., the amount of effort in collecting, sanitizing and treasuring the data).  These aspects, which we will not discuss in this paper due to space limit, are surveyed e.g., in \cite{oreilly2020,oreilly2021}.
 
\fakepar{\landingai} The above business considerations are relevant for the networking domain too. While this paper mostly considers AI from a technical viewpoint, first discussing how AI fits in the network architecture (Sec.\ref{sec:adn}) and next introducing examples of AI usage in current networks (Sec.\ref{sec:current}), it will also devote space to discuss legal (Sec.\ref{sec:future:aiact}) and human interaction (Sec.\ref{sec:future:xai}) aspects, that are particularly relevant for successful application of AI to networking.

\begin{comment}
Right now, the success factor of AI is that it  matched or exceeded human abilities  in  NLP, image and game.  
At the same time, AI consumed a significantly large amount of power to do so.
Moreover, the human brain comprises \DR{LOTS Of}neuron and synapses for an estimated 20W
of power.  Discoveries that biological neurons are actually much more complex (A 5-8 layers artificial deep neural network  is necessary to model a single human neuron \cite{beniaguev2021neuron}.  
If it's not green... right now it's beyond moore law.
While Green Network\cite{greenNet} is in the 2010, 
the emergence of   Green AI has been more recent \cite{greenAI}.
As such, future advances are expect to bring AI to an economically 
interesting operational standpoint (e.g., binary networks, XOR networsk, spiking networks).
\end{comment}

%% file: 30_adn.tex
While the specific domain of application of AI in networks are very diverse (e.g., from fiber/WiFi/5G/6G access, from campus to WAN and data-center networks to name a few)  we can identify commonalities across all these different environment. Indeed, while the resources to manage and the goal of their management  differ across the environments,  we argue that the underlying family of AI algorithms to empower the current and future generation of networks and protocols are similar, can be fit in a single  unified architecture.  Our aim in this paper is not to present  fully-fledged architectural details, which is the goals of standardization fora\footnote{Relevant fora are e.g., ETSI Experiential Network Intelligence\cite{etsi-eni}\censure{with a focus on AI}, ETSI Zero-Touch\cite{etsi-0touch}\censure{on autonomicity} and IRTF Computing in the Network (COIN)\cite{ietf-coin}.\censure{Research Group on programmability required to support emerging  edge network analytics, machine learning and deep learning workloads}}. Rather, this sections aims at introducing the ``DNA of the ADN'', i.e., the broad \emph{architectural principles} and the key hardware and software \emph{building blocks} of the autonomous driving network.

%COIN
%shift from data center (DC) toward edge computing and will debate whether this shift can be viewed as a cloud continuum. COIN specifically will focus on the evolution necessary for networking to move beyond packet interception as the basis of network computation. While existing DCs employ rudimentary languages for programming switch, richer programmability is required to support emerging workloads, such as edge network analytics, machine learning and deep learn. S

\subsection{Three-tiered Network AI Architecture}
\begin{figure}
    \centering
    \includegraphics[width=\columnwidth]{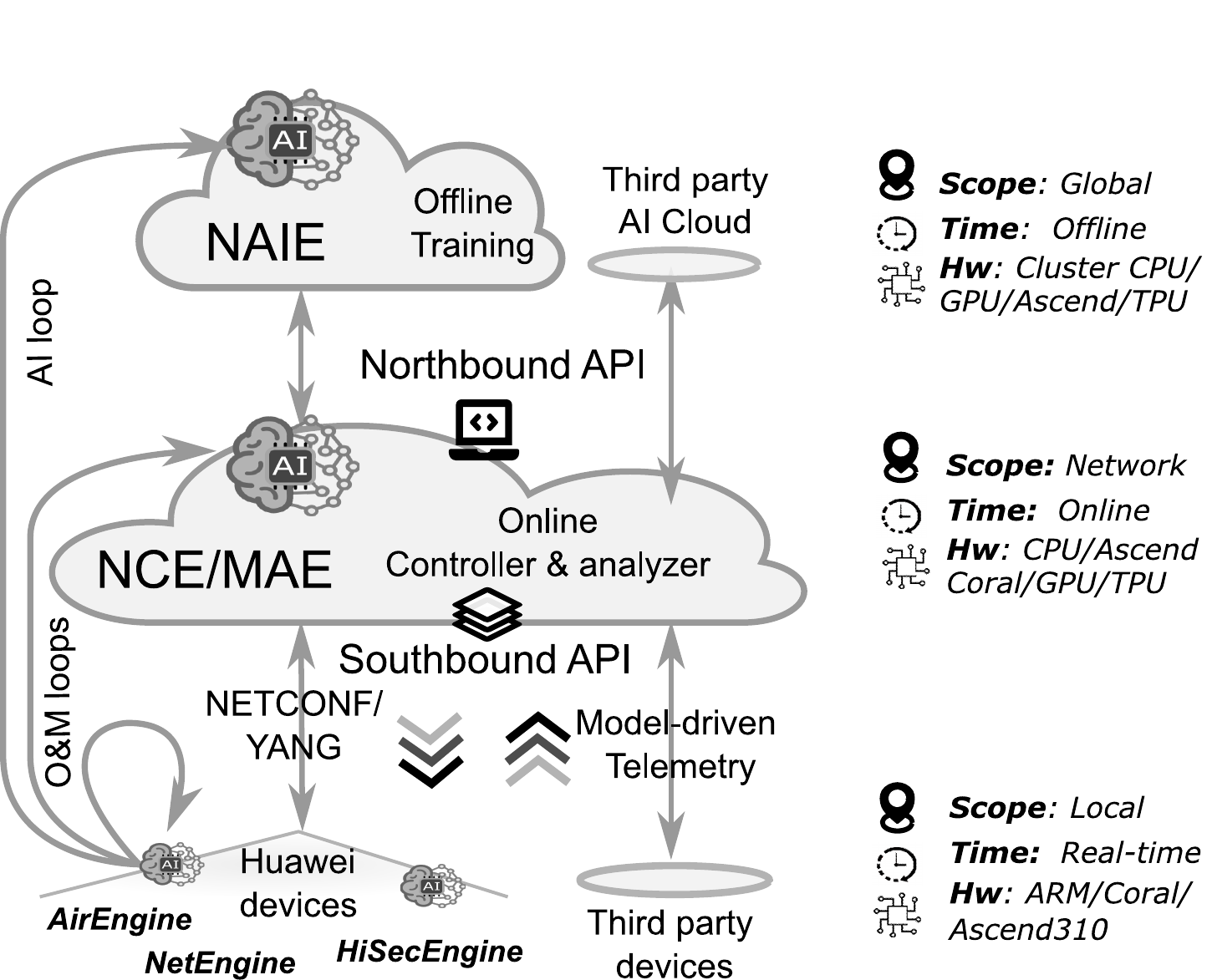}
    \caption{\textit{Network AI architecture}: three-tiered logical structure to fit different AI tasks from fast and local (device), to interactive and network-level (fog/cloud), to offline and global (cloud). The tiers interact in two kinds of closed loops for O\&M and AI operations (see Sec.~\ref{sec:adn:loop}).}.
    \label{fig:arch}
\end{figure}

The ADN comprises several physical elements, arranged from a logical point of view in a three-tiered architecture: this stem from the fact that  spatial properties of the network, and the time constraints for its management, require a distributed and hierarchical organization of the AI functions. 
Spatio-temporal aspects are reflected in the synoptic illustrated in Fig.~\ref{fig:arch}: the lowest tier is represented by 
fast and local AI decisions,  whereas online AI actions requiring a network-wide knowledge fit the middle tier, and finally offline AI tasks with global multi-network significance fit the  top tier.  Open APIs are necessary for  northbound interfaces, to interoperate with third party cloud platforms and AI software stacks: the top two levels of the architecture are also instrumental into automatically converting ``intent'' into configuration actions. Open APIs are also  needed for  southbound interfaces,  to accommodate third party devices: the bottom levels of the architecture interact  by  upstreaming measurement from  devices, as well as downstreaming  configuration actions, in a closed ADN control loop.

\fakepar{Device AI}
Local decisions that have to be taken in near-real time are delegated close to or onboard of devices (e.g.,  Huawei's
Net/Air/HiSec\,Engines~\cite{netengine,airengine,hisecengine}). For instance, augmented ``visibility'' tasks that require per-packet or per-flow decisions (scheduling, shaping, application identification, QoE estimation, anomaly detection) can be run at the edge. Devices  (e.g., WLAN AP or datacenter switches) can take autonomous decisions (e.g., auto-configuration and tuning) based on their local knowledge.  Such tasks might benefit from  low-power hardware acceleration chipsets for AI tasks (recall Tab.\ref{tab:tpu}) or just be equipped with standard CPU and benefit from software acceleration, so that  bespoke and highly-specific AI software stacks are expected to be the norm.

Further, non-technical considerations can favor in-device processing (e.g.,  business-sensitive or privacy-related for  GDPR compliance) to keep data local to CPE vs processing data remotely. 
Other considerations, related to hardware availability or processing vs bandwidth cost (e.g., deploying many weak in-device accelerators vs  fewer fog-accelerators or even fewer but more powerful cloud devices) can also affect the suitability of local vs remote processing for specific tasks.

\fakepar{Online fog/cloud AI}
Decisions that require network-level knowledge, and that do not have strict sub-second latency requirements,  such as for  controller and analyzer tasks, can be offloaded to the fog/cloud (such as Huawei's iMaster NCE/MAE~\cite{imasternce,imastermae} for the fixed/mobile network segments respectively). Controller actions can complement (or substitute) the one taken by devices, such as adding a slower centralized intelligence (e.g., taken by a single WLAN AP controller) on top of fast distributed  decisions (e.g., taken by several WLAN APs). Analysis tasks can provide a broader knowledge than that locally accessible, by e.g., correlating anomalous events at network scale for troubleshooting.  

From a hardware viewpoint, fog/cloud resources amortize Capex investments related to hardware acceleration, so that GPU and TPU should be expected to be more easily available whereas from a software perspective, code can leverage common and popular AI stacks. Controllers can access 
multi-vendor devices using cross-vendor southbound APIs (e.g., NetConf/Yang) to both upstream model-driven telemetry to feed AI decisions, as well as downstream automated configuration decisions to devices.
From a business viewpoint, the fog/cloud AI can be offered as a service, which tradeoffs Capex investments for higher bandwidth usage.

\fakepar{Offline cloud AI}
Knowledge that goes beyond the operation and management of a single network, or that span a large timescale, is better fit for offline storage and processing in the cloud (such as Huawei's iMaster NAIE~\cite{imasternaie}). This includes for instance model catalog management, model training services, transfer learning, federated learning, model health tracking, etc. These actions are instrumental for the good operation of network AI models, but are infrequent or anyway hardly need to be performed in a dynamic and interactive fashion (e.g., even if models are updated daily, data collection can be continuous, while training can be done nightly). As the offline AI cloud itself can become a bottleneck, research on the complementary  ``network for AI'' viewpoint tackles the optimization of AI workloads from a network system perspective (see for instance~\cite{lao2021atp,fei2021efficient} and references therein).

From a hardware viewpoint, these AI tasks can leverage a large fleet of cloud resources, spanning several type of AI accelerators. From a software viewpoint, the offline cloud can offer managed services (e.g., Huawei's   ModelArts\cite{modelarts}) but remains open and compatible with 
alternatives stacks (e.g., Amazon SageMaker\cite{sagemaker}) and model
marketplaces (e.g., Acumos\cite{acumos}),  further supporting open-source development through open APIs.

\subsection{Network AI and  autonomy levels}\label{sec:adn:loop}
With reference to autonomous driving vehicles, the industry identifies 
several levels of increased autonomy, from driver assisted (L1)  to partial (L2), conditional (L3), high (L4) and full (L5) automation. A similar categorization has then be extended to the context of network automation:  the goal of the ADN is to transition toward increasingly autonomous loops, where frequent human intervention (e.g., for technical necessity) is gradually substituted with sporadic human supervision (e.g., for legal aspects).   

In the path that leads to a fully autonomous network, we can identify and map the needed AI and ML techniques to reach a given level. In simple terms, we differentiate between techniques that need to be applied in \emph{open-loop} at that can greatly assist in augmenting the knowledge about the network operation (L2-3), as well as techniques applied in \emph{closed-loop}  to increasingly actuate or learn from the network (L4-5). 
We now briefly describe these simplified categories from the viewpoints of O\&M and AI loops illustrated in Fig.~\ref{fig:arch}.

\fakepar{O\&M loop}
An important aspect is to consider if the AI-enabled O\&M building block is working on open-vs closed-loop mode from a networking perspective. 
Supervised/unsupervised AI techniques are fit for \emph{open-loop O\&M}  tasks such as application identification, traffic and quality forecast, imputation for missing data, compressed/enhanced telemetry, fault and anomaly detection etc.  It should be clear that such techniques  mainly needs data to be fueled (and possibly labels for supervised learning) and can be   either operated in open-loop (sufficient for L2-3) or closed-loop AI modes (necessary at L4-5). These building blocks are fit for application on devices (or at the edge, depending on the timeliness vs computation requirements), although some tasks require the cloud processing power (for training or big data analytics).

\emph{Closed-loop O\&M}  can leverage AI techniques for several tasks, ranging from resource allocation, configuration adaptation, fault prevention and repair.  O\&M loop can be fully distributed, or have a  termination point in the fog/cloud, where centralized decisions can complement distributed decision.  As learning directly from the real deployment can be hazardous (performance during a cold-start learning phases will be bad),  it is desirable to pre-train in an actionable controlled environment (e.g., simulator, emulator) before further refine learning (e.g., digital twins,  real network).  AI techniques for closed-loop O\&M are intrinsically operated in closed-loop AI mode, although the O\&M loop can be closed in different
network architectural points (i.e., device/edge/fog/cloud) depending on the specifics constraints of the application use-case (e.g., latency, telemetry, timescale, processing power, law, etc.).

\fakepar{AI loop}
We refer to \emph{open-loop AI} to models that do not evolve (e.g., inference of a trained supervised model for regression,  classification or actuation) or intelligence that does not  trigger further analytic (e.g.,  periodic batch-mode data mining over a data-lake). Open-loop AI may take part in the device (e.g., inference) or the cloud (e.g., transfer learning, federated learning). 
Open-loop AI techniques are necessary for L2-3 network automation, and will be instrumental also for L4-5.  

Conversely, we refer to  \emph{closed-loop AI} as to the fact of altering the AI models themselves: we point out that this can happen with any of the \emph{supervised}  (e.g., incrementally training a model due to behavior drift of existing classes or  appearance of new classes), \emph{unsupervised} (e.g., stream-mode algorithms that alter existing models at any new sample) and \emph{reinforcement}  (e.g., continuous exploration phase throughout the whole lifetime of a reinforcement learning, or  bandit models) learning classes.  Closed-loop AI techniques will be necessary to  reach  L4-5 network automation.

\subsection{Network AI Software}
Ultimately, the ADN network is executing AI functions which are implemented as software instructions. As any software, AI models need to be designed, maintained and upgraded: even for the simplest L2 O\&M task, closed-loop management of AI software is key to successful deployment, which we briefly discuss.  

\begin{figure}
    \centering
    \includegraphics[width=0.9\columnwidth]{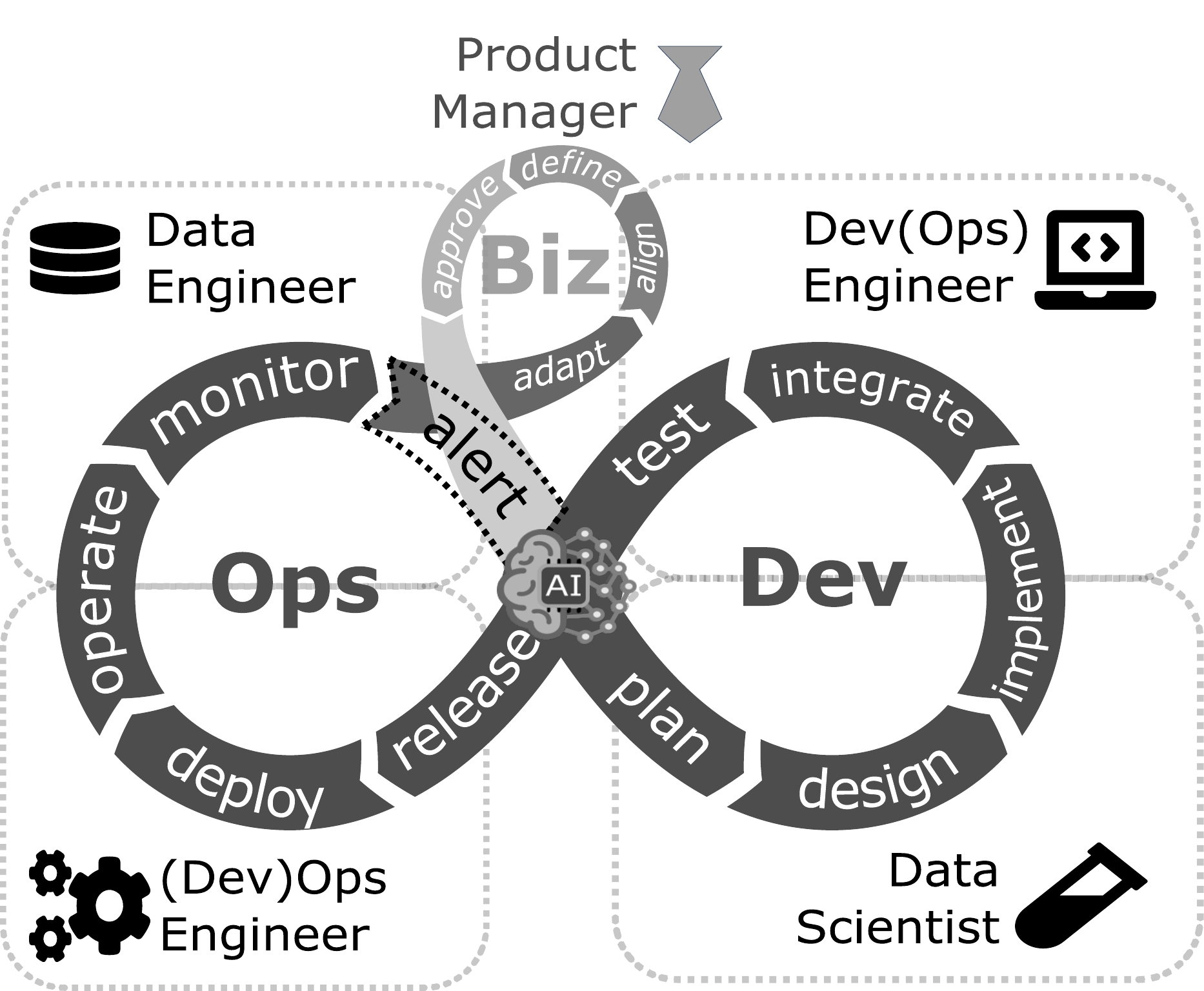}
    \caption{\textit{Network AI software management}: 
     main actors in the BizDevOps agile loop, with special focus on the AIOps roles and interactions (illustration adapted from  \cite{AIOPS-NETMAG21}).}
    \label{fig:aiops}
\end{figure}

\fakepar{AIOps Software Lifecycle}
AI software has a peculiar lifecycle, termed  AIOps  by Gartner, as  a particularization of the BizDevOps cycle to take into account specific characteristics of AI-software development. 
In a nutshell, DevOps is an agile methodology that combines software development (Dev) and IT operations (Ops), to shorten the systems development life cycle by providing continuous delivery, possibly further integrating Business (Biz) aspects. Complementary to the Dev and Ops engineering teams in classic IT and O\&M scenarios, AI development requires additional skillsets, which are identified in the Data Engineering and Data Scientist roles respectively, as illustrated in Fig.~\ref{fig:aiops} (further emerging roles are discussed in Sec.\ref{sec:future:data}).

%For instance,  several DL models are subject to a supervised training phase,  in which the model is presented with relevant (and abundant) examples of the objects to recognize (classification) or the real-valued function to learn (regression). With respect to AIOps, understanding if and when deployed DL models need to be retrained is thus an important task. 

\fakepar{AI Software in the ADN}
Taking the supervised case as an example and with reference to Fig.\ref{fig:aiops},  data scientists leverage data gathered by data engineers for model design (e.g., choosing the appropriate ML/AI family,   properly train by avoiding overfitting  and biases,  hyperparameter tuning, etc.). As data science skills may not accessible to all companies, to lower the startup costs, a recent trend is to automate part of the data scientists ML task (e.g., AutoML\cite{nas-survey}). Alternatively, the existence of open APIs makes it possible for the emergence of AI marketplaces (e.g., Acumos\cite{acumos}), that can offer readily trained models, which may lessen the need of data scientists for the most common tasks.
With respect to the  three-tiered network architecture early illustrated in  Fig.\ref{fig:arch},  models gathered by any of the above means can be deployed in edge devices or online flog/cloud (depending on capabilities of the devices).

%(autoML) procedures: \DR{for instance, Network Architecture Search (NAS). can simplify the model design part to cope with lack of data science expertise.} 

After models are properly trained and tested, they are deployed by Ops engineer in production system. Data engineers then help monitor data from deployed DL models: tracking of model data is then used to adapt, align and define the priorities (BizDevOps loop) or to simply alert on necessary updates to the model (DevOps loop) for the next Dev phase.  As early outlined, due to data drift, or environmental changes, deployed models may no longer be fit to the environment in which they are deployed, and shall need retraining. 
In the three-tiered ADN architecture, the offline AI cloud  is   responsible for tasks such as data lake storage and model catalog management, including model training and maintenance (eg., pruning to fit devices with constrained capabilities,  or adaptation in case of  unsupported operators on a device, etc.).

\begin{figure*}
    \centering
    \includegraphics[width=\textwidth]{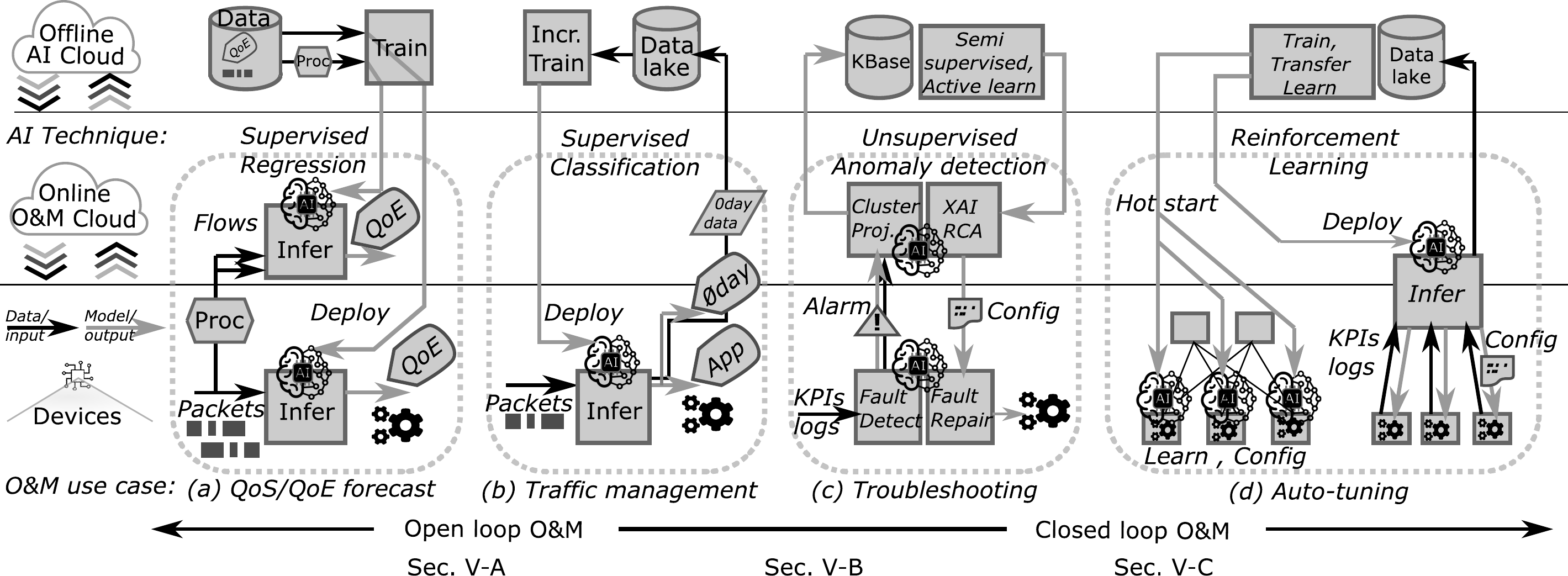}
    \caption{Synoptic of selected examples in current achievements of AI-assisted networking use cases of: (a) QoS/QoE forecast, (b) traffic management, (c) troubleshooting and (d) auto-tuning.}
    \label{fig:current}
\end{figure*}

%% file: 40_current.tex
We now make practical examples of AI-assisted network O\&M, that we map over the ADN architecture and illustrate at a glance in Fig.\ref{fig:current}, ranging from (a) forecast, to (b) traffic management, (c) troubleshooting and (d) auto-tuning. While not exhaustively covering the set of network applications and machine learning techniques\footnote{For instance, we plan to address the use of natural language processing techniques for the purpose of network configuration in a separate article.}, the selected set of examples fully span the whole set of supervised, unsupervised, semi-supervised and reinforcement learning branches overviewed in Sec.\ref{sec:ai}, in furthermore open and closed loop modes for both AI and O\&M viewpoints.

In this vision paper, we prefer to keep  discussion at a qualitative level: we thus avoid embedding  quantitative results that are already published elsewhere, to which we rather point the reader to.  Additionally, we provide insights from the real problems that AI can find in deployment, that the academic community may not be exposed in their day-to-day work, and is thus less sensitive to: in particular, in each use case we comment AI results under the angles of its (i)  generalization capabilities and (ii) benefits vs cost tradeoff -- that  Sec.~\ref{sec:cfr:data} and Sec.~\ref{sec:cfr:hwsw} respectively outlined being of key importance for successful transition from research to products.

\subsection{Efficiently handling the known (L1 to L2)}\label{sec:current:known}
Supervised ML techniques, such as regression and classification,  are apt at tackling well-specified problems in open-loop O\&M settings,  to increase visibility about network traffic or distill useful knowledge and information from raw data.  
In this section, we  outline both success and limits for two specific examples of application of each techniques.

\fakepar{Regression (e.g., QoS/QoE estimation)}
Regression techniques are fit for forecasting, e.g., future traffic demand or user behavior, or for learning complex  relationships, such as  relating network Quality of Service (QoS) indicator to user Quality of Experience (QoE) as exemplified in Fig.~\ref{fig:current}-(a).   In this latter context, a large body of literature employed ML techniques, to e.g., learn QoS indicators such as latency distribution \cite{deepq-netaiml20,routenet-sosr19} from topology, traffic matrix and routing information, or learn QoE indicators   for specific applications such as Web\cite{TNSM-20wikiqoe,TNSM-21-qoe}, video\cite{video-qoe1,video-qoe2} or games\cite{game-qoe}.

AI is desirable in this case, as it can leverage massive volume of data\footnote{It may be relevant for readers interested in the 
Web QoE specific use-case, that we made several datasets,  collected  in cooperation with Orange Labs~\cite{dataset-qoe-orange1,dataset-qoe-orange2} and Wikipedia~\cite{dataset-qoe-wiki} publicly available.} (e.g., automatically collected network~\cite{deepq-netaiml20,routenet-sosr19} or application~\cite{TNSM-21-qoe,video-qoe1,video-qoe2,game-qoe}  QoS/QoE indicators) without incurring high labeling cost (unless human opinion is explicitly factored in, as in \cite{TNSM-20wikiqoe}). On the one hand, work such as \cite{deepq-netaiml20,routenet-sosr19,TNSM-20wikiqoe,TNSM-21-qoe,video-qoe1,video-qoe2,game-qoe} proves that data-driven models provide accurate solutions, at both packet or flow-levels.  On the other hand, generalization capabilities and cost bares additional discussion.

As far as generalization is concerned, it is clear that models are tested on a subset of the whole set of applications, games, Webpages, terminals and network conditions. However, real products will be exposed to such diversity: thus, particular attention is needed to stress-test model capabilities beyond the classic techniques (e.g., k-fold cross validation).
For instance, as done in \cite{TNSM-21-qoe} for the case of Web QoE, it is useful to systematically analyze model bias induced by training data, and provide guidance for extending data corpus to reduce such via incremental training at higher autonomy levels (L3 and beyond). 

Additionally, the academic community often strives to increase accuracy of the proposed solution, irrespective of its computational cost. Yet, as diminishing return has to be expected beyond a given accuracy, solutions that explicitly allow to tradeoff (slight) accuracy loss for (significant) computational savings are to be preferred~\cite{TNSM-21-qoe}: the first role of AI researchers is, after all, to judge whether AI is the right tool for the problem at hand -- i.e., to avoid that all problems seems nails once you have an (AI) hammer.

\fakepar{Classification (e.g., Traffic management)}
Traffic management of Fig.~\ref{fig:current}-(b) is another relevant example where open-loop AI techniques are helpful:  traffic prioritization needs coarse-grained traffic category labels,  while policing may additional require fine-grained application labels.

In this context, AI is clearly beneficial since encryption is mandating to phase out Deep Packet Inspection (DPI) for behavioral classifiers, with satisfactory accuracy performance often in excess of 95\%~\cite{aceto2018tma,TNSM-21tc}.  Labeled data is in this case well known to be harder to gather and share~\cite{moore2005pam},  although the process can be automated to some extent~\cite{aceto2019mirage}. As the application coverage of publicly available datasets, is smaller than commercial grade needs, the question about generalization capabilities is manifest: public datasets generally comprise 10-50 classes whereas commercial grade ones\cite{TNSM-21tc}  include 2000+ labels, with the top-200 classes covering about 95\% of the traffic  volume. To cope with this mismatch between data accessible to academic vs relevance for industrial needs, we are currently in the process of releasing a highly-anonymized version of our commercial-grade dataset~\cite{TNSM-21tc} as part of Huawei Rapid Analytics \& Model Prototyping (RAMP) data challenges~\cite{xianti}.

As for the complexity is concerned, we observe that system researchers~\cite{XiongZ19,abs-1909-05680,swamy2020taurus,abs-2009-02353} consider extremely simple models (with just 21 neurons \cite{swamy2020taurus} or 50  neurons \cite{abs-2009-02353} \emph{overall}),  whereas AI researchers train excessively big models (state of the art models compared in~\cite{aceto2018tma} employ in excess of  hundreds-thousands neurons \emph{per-class}).  Awareness of commercial-grade challenges and constraints helps
landing commercial-grade models out of the lab, by explicitly parsimonious AI-model design (less than hundred thousands neurons for all 200 classes\cite{TNSM-21tc})   and optimized implementation (e.g., using  domain specific accelerator and languages~\cite{SEC-21,SIGCOMM-20}).

% In more detail, classic ML models are adapted  e.g., in \cite{XiongZ19}  to the match-action table model in P4, while \cite{abs-1909-05680} implement a random forest model to classify traffic. Deep NN models are instead used in \cite{swamy2020taurus} (with ASIC acceleration)  and \cite{abs-2009-02353} (with SmartNIC acceleration). These models are extremely simple and are \emph{orders of magnitude} smaller than those produced by AI  literature  on traffic classification.

\subsection{Taming the unknown (L2 to L3)}
As Jean Piaget famously said, ``Intelligence is not what you know, but what you do when you don’t know''. With this regard, supervised models are however inherently limited,  as they are e.g., unable to guess QoE or labels of completely new applications. Awareness of supervised AI limits is the first step to move up in the autonomy level.
%\DR{Intelligence is not what you know, but what you do when you don’t know (Jean Piaget)}.  

\fakepar{Anomaly detection (e.g., troubleshooting)}
Troubleshooting of Fig.\ref{fig:current}-(c) is an example use-case where supervised techniques are  not be a good fit. First, as networks strive to operate at very high   reliability  (i.e., 5-nines),  anomalies are rare events (so by definition only very few examples might be  available for training). Second, heterogeneity on the collected data and across different networks environments make generalization even harder (and thus unsupervised techniques appealing).

Clearly,  the use of lightweight unsupervised algorithms~\cite{ICDM-20,TNSM-20b} or self-supervised neural network~\cite{lstm2016icml,xu2018vae_unsupervised}  guarantees generalization by design: given the time-varying multi-variate nature of KPI data, both batch-mode~\cite{ICDM-20} (e.g., for periodic analysis) and stream-mode~\cite{TNSM-20b,xu2018vae_unsupervised} (e.g., for continuous analysis and trigger) algorithms can be leveraged, with stream-mode preferable due to the nature of the application.  However, no single anomaly detection algorithm can fully solve the entire end-to-end  troubleshooting pipeline~\cite{flap2017kdd}, which includes further steps such anomaly aggregation across multiple devices (requiring network visibility and thus a better fit for the private O\&M cloud). This algorithmic split across device/cloud gives the opportunity to fine-tune algorithmic results with, e.g., semi-supervised XAI~\cite{ITC-20} or causal RCA~\cite{pearl2019cacm} approaches, to ameliorate accuracy by exploiting previous information that might be available only for anomalies that occur more often (and that can benefit for  global-level knowledge in the cross-network AI cloud).

Along the complexity angle, unless  DL models (such as  recurrent LSTM~\cite{lstm2016icml},  VAE~\cite{xu2018vae_unsupervised}) are used for the anomaly detection task,  unsupervised techniques are generally lightweight. Yet, it is important to observe that the relative complexity among unsupervised algorithm still spans several orders of magnitude~\cite{TNSM-20b,arXiv-ad}, and that devices may be equipped with low computational power (recall Tab.\ref{tab:tpu}).
Particularly, as telemetry bandwidth is a more stringent  bottleneck in this case, algorithms have to consequently be deployed on edge devices for local execution (so that only part of the telemetry is then pushed to
online cloud for network-level analysis and visibility),  for which saving CPU cycles for similar algorithmic performance is still highly relevant.

\fakepar{Out-of-distribution detection (e.g., traffic management)}
Supervised technologies remains a good fit for some use-cases, as for instance in the early introduced traffic management of Fig.\ref{fig:current}-(b): yet,  models have a limited knowledge, as they are likely trained with only a fraction of the existing applications.  Applications that have never been presented to the model at training are called ``out of distribution'' (OOD) in AI terms, or ``zero-day'' (0D) in O\&M terms: when presented with OO/0D instances, any supervised model would misclassify  them as one of the known classes it has been presented during training.  The ability to detect OO/0D samples illustrated  in Fig.\ref{fig:current}-(b)  sits as an intermediate step~\cite{AIOPS-NETMAG21}, between \emph{open loop} (L1/L2, train once and deploy forever) and  \emph{closed loop} operation (L4/L5, to continuously train and deploy models). %as described in Fig.\ref{sec:adn:loop}).

By design, complementing supervised models with OO/0D capabilities increase their generality.
Both the AI and the O\&M communities have come up with general~\cite{NDsurvey} or use-case specific ways~\cite{zhang15unknown-ton} to deal with the problem.
We point out that for DL architectures, we have contributed a very effective  gradient-based technique~\cite{TNSM-21tc} that does not require architectural changes and works on unmodified models.

At the same time, it is worth reminding that this OO/0D additional feature comes at additional cost: in particular, 
while the gradient-based technique is faster than feed-forward computation,  some of the techniques we experimented with are  significantly slower than the DL inference itself~\cite{TNSM-21tc}, and as such are not practical, or need to be used sparingly~\cite{AIOPS-NETMAG21}. 
Furthermore, whereas OO/0D is useful in solving part of supervised models limits, further effort is needed to  assist~\cite{zhang20infocom-unknown,van2020flowprint} and explain~\cite{INFOCOM-20pandorabox} automated labeling of OO/0D traffic. Additionally, since few OO/0D  samples will be initially available, additional techniques (e.g., few-shots~\cite{few-shot-survey}   and class-incremental~\cite{TMA-21} learning) will be necessary to close the learning loop.
 
\subsection{Learning to learn (L4 and beyond)}
Higher levels of automation imply the use of closed-loop AI techniques in closed-loop O\&M settings. As Einstein famously said, ``\emph{The true sign of intelligence is not knowledge but imagination}'', and to achieve automation at L4 and beyond, the ability to continuously and efficiently learn is key.  

%We not only consider the problem of model deployment, but 
%additionally consider the problem of lifelong learning and model adaptation.

\fakepar{Reinforcement learning (e.g., auto-tuning)}
A classic example of closed-loop O\&M is represented by  automating resource usage and control, with either centralized fog/cloud agents (as in the WLAN use-case exemplified in the  right part of \mbox{Fig.~\ref{fig:current}-d)},  or distributed  agents on devices (as in  DCN case shown in the left part of \mbox{Fig.\ref{fig:current}-d)}. Note that the choice of distributed vs centralized intelligence may depend on timescale (e.g., DCN)  or other consideration (e.g., WLAN Campus AP controller vs individual home APs), thus the examples of  Fig.\ref{fig:current}-(d) are not meant to  exhaustively cover  all valid possibilities. 

In this context, AI agents are employed to reach an objective related to QoS (reducing flow completion time in DCN~\cite{ecn2021sigcomm}, improving signal quality in  WLAN~\cite{NETWORKING-21,AAAI-22}) or QoE (e.g. of videos~\cite{pytheas} and games~\cite{INFOCOM-20gameqoe}). To attain such goal, agents receive a reward as a result their action (e.g., setting threshold for ECN marking~\cite{ecn2021sigcomm}, WLAN channel~\cite{NETWORKING-21} or power~\cite{AAAI-22}  configurations, CDN node selection~\cite{pytheas}, or relative priority of game traffic~\cite{INFOCOM-20gameqoe}). In all these disparate cases cases,  AI is used to guide the exploration of an otherwise very large state space: e.g., from simpler Stochastic Bandits used in~\cite{pytheas}, to more complex Deep Reinforcement Learning (DRL) in~\cite{ecn2021sigcomm,NETWORKING-21}, or Transformers in~\cite{AAAI-22}. 

Generalization capabilities are important yet hard to ensure, as the environment in which an agent has been trained may differ significantly from the environment where it is deployed. As a rule of thumb,  frequent actions give more opportunities to learn: e.g., ECN threshold setting happens on a DCN RTT timescale~\cite{ecn2021sigcomm}, whereas CDN node selection happens on a per-session basis~\cite{pytheas}  and  WLAN AP configuration on a hourly basis~\cite{NETWORKING-21}.
Additionally, even though algorithms may account for online learning~\cite{ecn2021sigcomm,pytheas}, they need to be seeded with an offline training phase~\cite{ecn2021sigcomm,NETWORKING-21}. In this offline phase, algorithms are trained with trace-driven approaches~\cite{ecn2021sigcomm} or via simulation~\cite{NETWORKING-21} for hot-start: the more realistic the offline training environment, and the more diverse the environmental conditions explored, the better generalization capabilities can be expected in the real deployment.

In terms of computational complexity, inference is fortunately faster than training, which is expected to be computationally costly. Additionally, in the case of offline training phase, often the bottleneck is represented by the cost of simulating the environment at each action step (even for DRL~\cite{NETWORKING-21} and transformer~\cite{AAAI-22} architectures), so that parallel execution is appealing.

%\item  closed loop WLAN \cite{NETWORKING-21} datacenter \cite{sigcomm21ecn}, game qoe  \cite{INFOCOM-20gameqoe}Clearly, it would be desirable to Generality \cite{AAAI-22} distributed ECN tuning in data center \cite{ecn2021sigcomm}

%To make examples where reinforcement learning is used to manage resources to optimize  QoE  \cite{pytheas} QoE-based management\cite{sigcomm} QoE is also important \cite{sigcomm} \DR{that manages to prioritize flows depending on their QoE level}.  Complementary to that, machines can be used to imitate human behavior, eg., to learn how to play so to use then AI bots to assess how their ability to play )e.g. score, kill/death ratio) can be used to tune/prioritize traffic among games.

\fakepar{Lifelong learning (e.g., automate model design and update)}
Finally,  a complementary viewpoint that encompasses all use-cases illustrated in Fig.\ref{fig:current} and is necessary to reach higher automation level is represented by closed-loop AI techniques. In ML what matters is the journey, not the destination: thus successful ADN deployments need to embrace the idea that any model will need to continuously evolve.
We illustrate here a number of reasons with the traffic management example of  Fig.\ref{fig:current}-(b) for simplicity.

Techniques under the  so called ``lifelong learning''\cite{lifelong-learning} umbrella are key for generalization. First, as new zero-day application will keep appear and old applications will be forgotten, there is need for incremental~\cite{TMA-21} and decremental~\cite{ cauwenberghs2001decremental} learning.  As existing applications will drift~\cite{carela2016streaming}, continuous learning will not necessarily only focus on adding new classes, but to update existing ones. As application behavior differ in heterogeneous environment, federated learning~\cite{fedavg,IJCAIFL-20,naie-fl} will additionally be needed for privacy or business-sensitive constraints. Lifelong model changes bring significant challenges in terms of learning due to the opposite curses of ``catastrophic forgetting'' (of previously learned information) vs ``intransigence'' (to learn new one). Meta-learning can help finding a general representation that assist solving the above problems~\cite{lifelong-learning}, but as early illustrated, data science skills may not be available at all steps in an organization. As a result, any closed-loop AI technique to automate  the data science workflow (e.g.,  to automatically search for the fittest neural  architecture~\cite{nas-survey} that is easily in/decrementally updated) is relevant in the quest for robust and lifelong generalization of AI models for the ADN.

 % Complexity is of course the hidden elephant in the room: it is sufficient to observe that the ``Green networking''~\cite{greenNet2010} predates by over a decade the corresponding ``Green AI'' wave~\cite{greenAI2020}, as we further discuss in Sec.\ref{sec:greenai}.

%\item webqoe gathering  and understanding real user labels: \cite{WWW-19wikiqoe},\cite{TNSM-20wikiqoe}, \cite{MedComNet-20} 

%item webqoe changepoint detection \cite{CNSM-20b}, 

%% file: 50_future.tex
While the previous section has shown that AI techniques can be profitably used to solve real problems and transferred to real products, however much remains to be done before  highly autonomous L5 O\&M operations  can be attained by the  ADN: this section discusses several challenges that need to be solved along the way.

%  \DR{with a bias toward networking, while ML/AI advances in other fields than networking, we point out a biased viewpoint than networking, and
% additionally emphasize aspects that the general AI community is less interested with}.

\subsection{Dirty data}\label{sec:future:data}
First, it is well accepted that there is no AI without data~\cite{noaiwothoutdata2011cacm} -- so a few points 
are worth stressing concerning data access, representation and goverance.

\fakepar{Data access}   While, as earlier introduced, networking data is more fragmented and heterogeneous than natural language or images, it is true that the community-wide effort to share dataset equivalents to image\cite{imagenet} or natural language~\cite{commoncrawl} lags far behind. Challenges~\cite{itu-challenge2020,itu-challenge2021,xianti} are a partial answer to this  problem, but the community should recognize the need to federate data collection efforts, as opposite to scatter them along many tiny challenges. Eventually,  complementary to marketplaces for AI models~\cite{acumos}, the emergence of data marketplaces~\cite{cacm2018datamarketplace}, clearly abiding to local and international laws as discussed next, is an interesting opportunity to cope with this problem. 

\fakepar{Data representation} Access to data only solves part of the problem, since is commonly accepted that 50-80\% of AI scientist work is spent on data preparation~\cite{80data},
and indeed ``lack of data or data quality issues'' is the first technical\footnote{After the top-3 reasons (``company culture'', the ``difficulty in identify business use cases'', and the ``lack of data scientists roles'') which are of non-technical nature.} bottleneck to AI adoption, as identified by respondents  (5\% of which are from the telecommunication and networking industry) to the Oreilly radar survey~\cite{oreilly2020}.  Whereas some amount of human assistance to AI models is needed, our goal is not to enslave human labor to reach AI success~\cite{anatomyofai,captchamadness}, or at least to consciously use as little human help as possible.

On the one hand, the  AI community has come up with a number of best practices and techniques to cope with data quality\cite{dirtydata} or lack of labels 
(via active-learning\cite{active-learning}, self-attention\cite{selfattention}, few-shot\cite{few-shot-survey} or  self-supervision\cite{selfsupervised}). On the other hand, data in network is intrinsically highly dimensional, topologically complex (several time-varying multi-layer logical graphs),  heterogeneous and multi-modal (logs, packets, timeseries, configuration, etc.) and misses a single, unified and universally accepted representation, which remains an open problem.

\begin{comment}
\DR{ 
To cope with data format: embedding
To cope with lack of labels:   Self-attention\cite{x}, and self-supervised learning \cite{xy}
To cope with few labels: Few-shot learning. Recognizing that the cost of labeling exceeds the cost of training, few-shot \cite{few-shot-survey}
 To cope with rare events: Fully unsupervised  techniques.\cite{xy}
 To cope with portability: transfer learning, few-shot or meta-learning 
}
\end{comment}

\fakepar{Data governance}  Finally, unless data, and meta-data are treasured at (or beyond) the level of first-class citizen for network AI, then much of the AI effort will likely end up being in the 80\% of failing projects~\cite{80fail}. The complexity of meta-data management and of properly granting access to data, calls for a  more systematic approach to data governance~\cite{abraham2019governance}: e.g., by the introduction of data stewards, beyond the roles of data owner, data engineer and data scientist.
Data stewards should govern the access to data, on behalf of data owners and using the process developed by data engineers, in compliance with regulation aspects.

\subsection{AI Regulation}\label{sec:future:aiact}
Second, AI will not be successful if it's not legally compliant.
As  the regulation ecosystem is fragmented as it differ from country to country\footnote{For instance, EU General Data Protection Regulation (GDPR)\cite{gdpr},
or the more complex maze of laws across US states\cite{uslaws}.}  and additionally evolves over time\footnote{As for the new 
proposal for a regulatory framework on Artificial Intelligence (AIAct)\cite{aiact} launched Apr 2021 in EU, or the  Personal Information Protection Law (PIPL)\cite{pipl} introduced in Nov 2021 in China}, we briefly review the expected impact on data and AI models.

\fakepar{Impact on data}  First, we observe that networking O\&M area is intrinsecally less sensitive with respect to other industries that treats biometric,  physiological or medical data (at the highest risk level for AI Act). Still, even though  most content is end-to-end encrypted\cite{https2017usenix}, it is useful reminding that even IP addresses are personal data according to GDPR, and have to be processed accordingly.
While some use-cases may raise limited risk,  however this is not going to be the general case: as such, lawful compliance of AI data processing should come as a fundamental architectural feature, and not as an afterthought.

For instance, e.g., security protection from intruders may process IP addresses of attackers received by Darknets\cite{CONEXT-21},  but may also include part of so called  ``backscatter''\cite{backscatter} traffic from compromised machines owned by unconscious citizens/companies. As of now, it is safer when data is not shared and stays under same-country (or same-regulation) boundaries: however evolving and heterogeneous regulation between countries  can clearly become a headache for AI researchers, and slow down AI adoption for global companies. The early introduced data steward role should additionally be coupled with a legal expert role, clearly mapping data to a specific risk category depending on its processing.  Data should also be tainted, so that crossing legal borders should be either lawful or impossible by design in the architecture.

\fakepar{Impact on models}  In addition to having an  impact on data sharing,  regulation impacts algorithms even when data is \emph{not} shared: as, for instance, in the case of federated learning~\cite{fedavg}.  Whereas loads of work ensure privacy of that federated learning process~\cite{truex2019flrprivacy}, however attacks~\cite{wang2019flprivacyleak} are still possible that yield to serious privacy leaks -- with possible legal consequences, when clients participating in the learning span across multiple countries and regulations.

Moving one step further, the ``right to be forgotten'' for privacy or business reasons raise the need for \emph{decremental} learning, which has recently enjoyed a number of proposal for Support Vector~\cite{cauwenberghs2001decremental}, Random Forest\cite{brophy2021machinenlearning_forest} or Neural Networks\cite{neel2020machineunlearning_gradient}. However, as pointed out in  \cite{chen2021machineunlearning_privacy} these algorithms are not immune to attacks, so that machine unlearning can have counterproductive effects on privacy. Further,  whereas probabilistic~\cite{sommer2020machineunlearning_probverif} may be  enough from a scientist point of view, it would strikes us as odd if this was considered acceptable from a legal standpoint. The lack of clear guidelines and definition can slow-down adoption, for which  non-trivial discussion between lawyers and scientists seems in order.

\subsection{eXplainable AI}\label{sec:future:xai}
Third, AI will not be successful if it's not trusted.
In the path to L5 ADN,  AI should free humans from the burden of the fast loop (that calls for automation), and facilitating the interaction with them in a slow loop (that calls for explainability).
In a sense, AI should smoothly transition from day-to-day technical supervision (i.e., as if AI is a junior colleague that needs coaching) to less and less frequent supervision  (i.e., AI becomes a senior respected colleague).

\fakepar{Fairness and generalization}
Trust can be gained only if AI can provide, in addition to good performance, also clear explanation of the decision process.  In other fields where ethical matters are prominent, fairness issues in AI models is well known and debated~\cite{googlefairnessissue}. Part of the lack of fairness is knowingly rooted in class imbalance in the data used in the first place \cite{fairnesssurvey}, for which countermeasures exist.  
If, in the  networking field, unfairness is less likely to have life-changing decisions, the existence of such bias can still severely affect the model generalization capabilities. Finally, it  has recently been shown that models can be accurate but for the wrong reasons~\cite{lapuschkin2019unmasking}, which can rapidly undermine the trust that AI technologies have gained so far.  
Trust between humans and machines~\cite{hoffman2013trust,wang2016trust} depends on understandability  and directability (how easy  one can assert, control or influence when something goes wrong): techniques to improve AI explainability~\cite{XAIsurvey} are therefore a key element in the future ADN.

\fakepar{Faithfulness and accountability}
Additionally, whereas most of AI decisions in network O\&M will not have dramatic life-changing impacts, such decisions should be accountable from a business and possibly legal (as per previously discussed regulation) perspectives. On the one hand,   probabilistic   \cite{sommer2020machineunlearning_probverif}  or post-hoc explanation of
surrogate models\cite{sigcomm20-xai,lundberg2017shap} may be sufficient for business-level accountability. On the other hand, proving law compliance at a forensic level  may require faithfulness,  a key XAI property defined as  correctly reflecting the system  process for generating the output \cite{phillips2020four}. 
This pushes to  embed explainability directly in the model, as recent work started advocating \cite{alvarez2018towards,rudin2019stop}.  We finally point out that while XAI may come with a cost (e.g., additional computation, or accuracy loss), this seems to be a necessary price to pay in light of the above issues.

\subsection{Green AI}\label{sec:future:greenai}
Fourth, AI revolution will not happen if it's not cost-effective -- as we previously observed, efficient
hardware accelerators were instrumental to AI success (Sec.\ref{sec:cfr:hwsw}), and cost-effective operation is a  precondition for business success (Sec.\ref{sec:cfr:business}).
Interestingly, we observe  that the ``Green networking''~\cite{greenNet2010} predates by over a decade the corresponding ``Green AI'' wave~\cite{greenAI2020}.  We examine here the complementary aspects of green AI from system and algorithmic perspectives.

\fakepar{Green system}
As earlier indicated, Gilder law forecasts that bandwidth scaling rate exceeds Moore law. Additionally, OpenAI estimated\cite{openai-compute} that the amount of compute used in the largest AI training runs in the last decade has increased exponentially, again faster than Moore law. As such, the carbon impact of AI\cite{ai2020carbon,wired-ai-burn-planet} has rightly come under scrutiny  (along with blockhain). 

With the example of the traffic management use case early introduced in Sec.\ref{sec:current:known}, we observe that commercial-grade models can run on DSA (e.g., Ascend310 TPU) supports inference rate at  100Gbps with a power drain of 7W.  However, we additionally point out that complementary techniques that avoids AI computation altogether (such as approximate-key-caching\cite{INFOCOM-22}) can further bring multiplicative speedups without compromising accuracy. In other technological fields, the awareness of fuel efficiency (for cars) or energy efficiency (for electrical appliances and even light bulbs) is strictly regulated and mandatorily exposed -- we believe that the same awareness should be extended to the AI sub-components of any networked system, by explicitly measuring  metrics such as \emph{task/Joule} or \emph{accuracy/Watt}.

\fakepar{Green models}
Fortunately on the other hand, the AI community has worked toward improving algorithmic efficiency -- so that, e.g., the number of floating point operations required to train a classifier to AlexNet-level performance on ImageNet has significantly decreased in the last decade~\cite{openai-efficiency}.

Always with reference to the traffic management use case of  Sec.\ref{sec:current:known}, recently introduced architectural innovations  such as  inverted residual as in MobileNet\cite{howard2018mobilenetv2} and  point-wise convolution as in ShuffleNet\cite{ma2018shufflenetv2}, significantly reduce model complexity and are amenable to be run on ARM even without any DSA acceleration\footnote{Currently unpublished results show
that our ShuffleNet re-implementation on TFLite\cite{tflite} achieves accuracy on par of \cite{TNSM-21tc} by taking 140$\mu{}s$ per classification on a single ARM Cortex A72 core with batch size $b=1$,  and is even faster than on NVIDIA P100 GPU with small batch $b=8$.}.   At the same time, AI model design is still an art, so that automated ML (autoML) techniques such as  Neural Architecture Search (NAS)~\cite{nas-survey} are appealing to assist AI model design. However, NAS is guided by an indirect measure of computation complexity (i.e., FLOPs). Therefore,  further research is needed to explicitly include in the NAS loop more direct metrics,  such as speed or energy consumption, to perform an \emph{ecology} of models design space (as well as  making the NAS process itself more energy efficient).

\subsection{AI-Native functions and architecture}\label{sec:future:ainative}
Finally, we need to point out that the  ADN architecture  illustrated and exemplified in this paper is a natural evolution of the current cloud-native network architecture. In other words, exactly as it happened for IP networks, where QoS, mobility, and security features have been added as afterthoughts to the original architecture, the same is now  happening for AI technologies.
We point out that we are not reviving here the debate of clean-slate vs evolutionary approach~\cite{cleanslate-vs-evolutionary} to network technology (r)evolution. Neither our goal is to anticipate the success of such changes~\cite{ammar2018exunopluria}, though we point out that time is ripe in terms of software stack and hardware support, which are known to have a disproportionate impact on technological success~\cite{hwlotteryCACM21,declarativeMLCACM22}.

Rather, we believe that this timeframe constitute  an opportunity for the network community to  rethink the design of an AI-Native  architecture, by holistically integrating AI in the whole network landscape, as opposite to just delegate specific tasks to AI islands.  In other words,  we believe that the future AI-Native architecture should allow to  expose, combine and orchestrate explainable AI functions, ``wired'' to data in a way that respects regulation, overall bringing improvements on network operation at a lower cost and energy footprint.

%
%as lego building blocs that can be deployed and are mutually aware of each other and of their impact 
%(e.g., if a scheduler in the middle of the network and a shaper at the edge attempt at doing conflicting objectives, one can see unwanted phenomena such as''reprioritization''\cite{REPRIO}). 
% Native AI: which functions should we replace by AI  ?
% Many things (QoS; mobility; security) were not added by design in the original network desgin.
% As we are re-architecting the Internet for
% security (SCION),    mobility (MobilityFirst), ZZZ, we can imagine --- if not a clean-slate, at least a concious redesign of the network to address which fundamental building blocks would be better replaced by AI.

%THis has consequences: Should we replace TCAMs with TPUs for packet classification \cite{NeuralPacketClassification}

%E.g., if we need DRL for MAC protocols, or WLAN alignment or inference at a packet basis, then
%we need to have hw support for that. Should that happen is also related to a tradeoff of 
%(i) the expected gain vs (ii) the expected cost.

%  IEEE  International Network Generations Roadmap (INRG)
% Artificial Intelligence
% and Machine Learning WG